\documentclass[prc,twocolumn,groupedaddress]{revtex4-1}

\usepackage{amsmath}
\usepackage{amssymb}
\usepackage{nicefrac}

\newcommand{\eref}[1]{(\ref{#1})}
\newcommand{\Eref}[2]{Eq.~\eref{#1}}
\newcommand{\half}{\frac{1}{2}}
\newcommand{\nhalf}{\nicefrac{1}{2}}
\newcommand{\grad}{\nabla}
\newcommand{\unitr}{\mathbf{\hat r}}
\newcommand{\unittheta}{\boldsymbol{\hat\theta}}
\newcommand{\unitphi}{\boldsymbol{\hat\phi}}

\begin{document}

\title{The effect of spin-orbit nuclear charge density corrections due to the anomalous magnetic moment on halonuclei}

\author{A.~Ong}
\email[]{ong.andrew@gmail.com}
\author{J. C.~Berengut}
\author{V. V.~Flambaum}
\affiliation{School of Physics, University of New South Wales, Sydney 2052, Australia}

\date{29 June 2010}

\begin{abstract}

In this paper we consider the contribution of the anomalous magnetic moments of protons and neutrons to the nuclear charge density. We show that the spin-orbit contribution to the mean-square charge radius, which has been neglected in recent nuclear calculations, can be important in light halonuclei. We estimate the size of the effect in helium, lithium, and beryllium nuclei. It is found that the spin-orbit contribution represents a $\sim 2\%$ correction to the charge density at the center of the $^7$Be nucleus. We derive a simple expression for the correction to the mean-square charge radius due to the spin-orbit term and find that in light halonuclei it may be larger than the Darwin-Foldy term and comparable to finite size corrections. A comparison of experimental and theoretical mean-square radii including the spin-orbit contribution is presented.
\end{abstract}


\maketitle

\section{Introduction}

It is known that the spin-orbit density of neutrons (and protons) contribute to the nuclear charge density as a relativistic effect~\cite{Bertozzi1972408}. Consider a reference frame in which there is no charge density present. By applying a Lorentz transformation, we see that the 0-$th$ component of the 4-current vector (the charge density) develops a non-zero value if the vector current is non-zero. The vector current exists in neutrons due to the anomalous magnetic moment.

The spin-orbit density effect of neutrons has been shown to contribute to the change in nuclear charge radius (and hence scattering cross-sections) between $^{40}$Ca and $^{48}$Ca~\cite{Bertozzi1972408} (see also~\cite{Kurasawa:2000ve} for more recent relativistic calculations). On the other hand the spin-orbit contribution was seen to have a negligible effect on the differential cross-section of $^{208}$Pb~\cite{Bertozzi1972408,PhysRevC.13.245}, and by extension other heavy nuclei.

The relativistic correction known as the Darwin-Foldy term~\cite{PhysRev.78.29, bjorkendrell1964} is by convention a recoil correction. In a recent review paper~\cite{PhysRevA.56.4579}, it was recommended that this correction be treated  as part of the charge radius, due to the fact that it appears in the charge density of the proton. This will be the approach we will adopt in our calculations.

Interest in the neutron contribution to nuclear charge densities has been renewed by the discovery of halonuclei. Comparison between theory and high-precision experiment has been performed in isotopes of He~\cite{PhysRevC.73.021302, Pieper:2007ax} and Li~\cite{ISI2001,PhysRevC.66.044310,PhysRevC.52.3013,PhysRevC.66.041302,Tomaselli2001298,PhysRevC.57.3119,PhysRevC.68.034305}. Relativistic corrections and finite size effects are consistently taken into account in these theoretical calculations, while the spin-orbit effect is never explicitly considered.

In this paper we calculate the spin-orbit contribution to nucleon charge density. We derive a simple, general expression for the contribution of the spin-orbit density to the mean-square charge radius. In light halonuclei this contribution may be comparable to the finite size effects, and may be larger than other relativistic corrections that are included such as the Darwin-Foldy term. In addition, we use the oscillator model to obtain numerical estimates of the spin-orbit charge density in $^7$Be.

We present the calculated values of the mean-square charge radii, including the spin-orbit contribution, in Table~\ref{tab:finalcomp}.

\section{Spin-orbit charge density in the nonrelativistic limit}

We begin with the expression for the 4-current derived from quantum field theory considerations~\cite{landau1982} (relativistic units $\hbar = c = 1$ are used throughout, unless otherwise stated),
\begin{equation}
\label{eq:4current}
j^\mu = \bar{\psi} \left( (4M^2 F_e - q^2 F_m)\frac{\gamma^\mu}{P^2} + \frac{2M}{P^2}(F_e - F_m)\sigma^{\mu\nu}q_{\mu\nu} ) \right) \psi
\end{equation}
where $M$ is the mass of the nucleon, $F_e$ and $F_m$ are electromagnetic form factors, $P^2 = 4M^2 - q^2$ is the momentum operator, $\mathbf{q}$ is the momentum transfer operator, and $\sigma^{\mu\nu} = \frac{1}{2}(\gamma^\mu\gamma^\nu - \gamma^\nu\gamma^\mu)$ is the antisymmetric tensor. For nucleons at the Fermi surface $\frac{q^2}{M^2} \sim \frac{p_F^2}{M^2} \sim 0.08$, therefore in this paper we perform all calculations in the non-relativistic approximation, accurate to order $\frac{q^2}{M^2}$. Here $p_F$ refers to the nuclear Fermi momentum.
In the non-relativistic limit the lower component of the Dirac wavefunction $\psi = \left(\begin{array}{c} \Phi\\ \chi \end{array} \right)$ is given simply by
\begin{equation}
\label{eq:nonrelpsi}
\chi = \frac{\boldsymbol{\sigma}\cdot\mathbf{p}}{2M} \Phi\ .
\end{equation}

The charge density can be calculated from the 0-$th$ component of the 4-current~\eref{eq:4current},\ which to order $\frac{q^2}{M^2}$ is
\begin{equation}
\label{eq:jzero}
j^0 = \bar{\psi} \left(\begin{array}{cc} F_e - \frac{q^2}{4M^2}(F_m - F_e) & -\frac{(F_e - F_m)(\boldsymbol{\sigma}\cdot \mathbf{q})}{2M}\\ -\frac{(F_e - F_m)(\boldsymbol{\sigma}\cdot \mathbf{q})}{2M} & - F_e + \frac{q^2}{4M^2}(F_m - F_e)\end{array}\right) \psi\ .
\end{equation}
The form factors $F_e$ and $F_m$ are, in general, functions of the momentum transfer squared $q^2$. However, for our purposes it suffices to consider only $F_e(0)$, the electric charge of the nucleon, and $F_m(0) - F_e(0) = \mu$, the anomalous magnetic moment ($\mu_n = -1.91$ and $\mu_p = 2.79 -1 = 1.79$ for neutrons and protons, respectively).

Substituting the nonrelativistic wavefunction into \eref{eq:jzero} gives the spin-orbit contribution to the charge density, to order $\frac{q^2}{M^2}$:
\begin{equation}
\label{eq:compactchargedensity}
\rho = \frac{\mu}{4M^2} 2i\,\boldsymbol{\sigma}_{\alpha\beta} \cdot \left[ \left( \nabla\Phi_\alpha^\dagger \right) \times \left(\nabla\Phi_\beta \right) \right]\ .
\end{equation}
Here there is an implied summation over $\alpha,\beta = 1,2$ --- the upper and lower components of the spinor $\Phi$. Note that the Darwin-Foldy term ($F_e\,q^2/8M^2$) can also be derived from \eref{eq:jzero}.

Calculation of the spin-orbit charge density is described in the Appendix.

\section{Charge density in spherically symmetric potentials}

Assuming a spherically symmetric nuclear potential, we can write the wavefunction in the shell-model as $\Phi = f(r)\,\Omega_{jlm}$, where $\Omega$ refers to the spherical harmonic spinors \eref{eq:omegadef}.
From~\eref{eq:compactchargedensity} we find that the anomalous charge density produced by $s$-waves is identically zero. This can be readily seen by noting that for each $s_{1/2}$ projection, the wavefunction only has either an upper or lower component, which means that summation over $\alpha$ and $\beta$ selects only the $\sigma_{z,\alpha\alpha}$ component (in the standard representation). Since the wavefunction components in this case are real, $\nabla \Phi^*_\alpha = \nabla \Phi_\alpha$, and therefore $\boldsymbol{\sigma}_{\alpha\beta} \cdot \nabla \Phi_\alpha^\dagger \times \nabla \Phi_\beta =  0$.

Consider what happens to the total local charge density when we sum over a spin-orbit doublet of nucleons. In the LS-coupling scheme, the $\boldsymbol{\sigma}_{\alpha\beta}$ term in Eq.~(\ref{eq:compactchargedensity}) results in the cancellation of $\rho$ for pairs of antiparallel spins. This implies that the sum over any spin-orbit doublet is zero. On the other hand in the JJ-coupling scheme there is no straightforward cancellation, and as a result the sum of the charge density contribution over a closed subshell is not zero. However, the sum over a spin-orbit doublet, for example $1p_{3/2}$ and $1p_{1/2}$, is zero if the radial wavefunctions are equal (any difference is of higher order in $\frac{q}{M}$).

For comparison with experiment, one wishes to know the contribution of the spin-orbit charge density to the mean-square charge radius. We define this for a single nucleon as
\begin{equation}
\label{eq:definitionofr2}
\langle r^2 \rangle_{so}^{(1)} = \int r^2 \rho\,d^3\mathbf{r}\ .
\end{equation}
Using this definition and \eref{eq:compactchargedensity}, we find
\begin{equation}
\label{eq:onenucleonr2correction}
\langle r^2 \rangle_{so}^{(1)} = -\frac{\mu}{M^2}(\kappa+1)
\end{equation}
where
\begin{equation}
\kappa = \begin{cases} l, & j=l-\frac{1}{2} \\ -(l+1), & j=l+\frac{1}{2} \end{cases}
\end{equation}
We defer proof to the Appendix. Interestingly, $\langle r^2 \rangle_{so}^{(1)}$ is independent of the radial wavefunction $f(r)$.

To compare calculated \textit{point-proton} mean-square charge radii with experiment, we introduce nuclear size and relativistic corrections~\cite{PhysRevA.56.4579} using the equation
\begin{equation}
\label{eq:rmschargeradius}
\langle r^2 \rangle_{ch} = \langle r^2 \rangle_{pp} + \langle R_p^2 \rangle + \frac{N}{Z}\langle R_n^2 \rangle + \frac{3}{4M^2} + \langle r^2 \rangle_{so}
\end{equation}
where
\begin{equation}
\label{eq:totalsorms}
\langle r^2 \rangle_{so} = \frac{1}{Z}\sum_i\langle r^2\rangle_{so}^{(i)} = - \frac{1}{Z}\sum_i \frac{\mu_i}{M^2}(\kappa_i+1)
\end{equation}
and $i$ runs over all nucleons (but there is zero contribution from a sum over spin-orbit doublets).
Here $\langle R_p^2 \rangle = 0.769(12)~\text{fm}^2$ and $\langle R_n^2 \rangle = -0.1161(22)~\text{fm}^2$~\cite{PDBook} are the mean-square charge radii of the proton and the neutron respectively, and the term $\frac{3}{4M^2} = 0.033~\text{fm}^2$ is known as the Darwin-Foldy term~\cite{PhysRevA.56.4579}. The effect we are considering simply adds the quantity $\langle r^2 \rangle_{so}$ to the equation used by previous works, e.g.~\cite{PhysRevLett.96.033002}. The factor $\frac{1}{Z}$ arises from normalization.

\section{Spin-orbit charge density in Beryllium-7}

As was shown in the previous section, $\langle r^2 \rangle_{so}$ does not depend on the radial wavefunctions. However, one may need to know the value of the charge density $\rho(r)$, which depends on the radial wavefunctions. To estimate the size of the effect, we will use the harmonic oscillator potential 
\begin{equation}
V(r) = \frac{M\omega^2r^2}{2}
\end{equation}
as an approximation to the nuclear potential. Using the known solutions of the quantum harmonic oscillator~\cite{griffiths2005},
\begin{align}
f_{s}(r) &= \frac{2^\frac{7}{4}\alpha^\frac{3}{4}}{\pi^\frac{1}{4}} e^{-\alpha r^2}\\
f_{p}(r) &= \frac{2^\frac{11}{4}\alpha^\frac{5}{4}}{\pi^\frac{1}{4}3^\frac{1}{2}} r e^{-\alpha r^2}
\end{align}
the form of the spin-orbit charge density $\rho_{jlm}$~\eref{eq:compactchargedensity} for the $l = 1$, $j = 3/2$ states can be written as follows
\begin{gather}
\label{eq:1p32big}
\rho_{\frac{3}{2}1\frac{3}{2}} = \rho_{\frac{3}{2}1-\frac{3}{2}}  =  -\frac{\mu}{M^2} A^2(r) (1 - 2\alpha(x^2 + y^2)) \\
\label{eq:1p32small}
\rho_{\frac{3}{2}1\frac{1}{2}} = \rho_{\frac{3}{2}1-\frac{1}{2}} = -\frac{\mu}{3M^2}A^2(r) (3 - 2\alpha(x^2 + y^2 +4z^2))
\end{gather}
where $A^2(r) = \frac{2^{\frac{5}{2}}\alpha^{\frac{5}{2}}}{\pi^{\frac{3}{2}}}e^{-2\alpha r^2}$ and $\alpha = \frac{M\omega}{2}$ parametrizes the potential. Note that the sum of these charge densities is spherically symmetric:
\begin{equation}
\sum_m \rho_{\frac{3}{2}lm} = -\frac{2A^2(r)\mu}{M^2}(2-\frac{8}{3}\alpha r^2) \ .
\end{equation}
$\alpha$ can be obtained through the fitting of the model to known experimental measurements.

To estimate the size of the spin-orbit charge density in $^7$Be, we use the shell model, where the neutrons and protons fill nuclear states independently. $^7$Be has a single unpaired neutron in the $1p_{3/2}$ state. The charge density generated by this neutron is given by \eref{eq:1p32big}. By comparison, in our model the usual charge density contribution from the four protons, taking into account only highest order terms, is ($e =1$ in our units)
\begin{eqnarray}
\label{eq:totalprotonchargedensity}
\rho_{p,total} &= & \sum_{i} \phi^{\dagger}_i \phi_i\nonumber\\
&=& \frac{2^{\frac{7}{2}}\alpha^{\frac{5}{2}}}{\pi^{\frac{3}{2}}}(x^2 + y^2 + \frac{1}{2\alpha})e^{-2\alpha r^2}\ .
\end{eqnarray}
The relative contribution of the spin-orbit charge density to the total nuclear charge density is the greatest at the center of the nucleus, where it represents a $\sim2\%$ correction. For this calculation, we used $\omega = \frac{40 MeV}{A^1/3}$ as our fitting parameter for the harmonic oscillator potential; this corresponds to $\alpha = 0.25$~fm$^{-2}$.

The effect of the spin-orbit charge density on the mean-square charge radius is given by \eref{eq:totalsorms}. For the neutron and two protons in the $1p_{3/2}$ state it is
\begin{align*}
\langle r^2 \rangle_{so}
 &= \frac{1}{Z}\frac{1}{M^2} (\mu_n + 2\mu_p) \\
 &= -0.021 + 0.039 = 0.02~\textrm{fm}^2
\end{align*}
This is a maximal estimate; configuration mixing reduces the size of the effect. Table \ref{tab:becomp} presents the experimental and theoretical charge radii of $^7$Be. It shows that the size of $\langle r^2 \rangle_{so}$ can be important, although it is smaller than the spread between different theoretical data points.

\begin{table}
\caption{\label{tab:becomp}Comparison of $^{7}$Be experimental and theoretical charge radii in fm$^2$.}
\begin{ruledtabular}
\begin{tabular}{ccccc}
&$\langle r^2 \rangle_{exp}$~\cite{PhysRevLett.102.062503} & $\langle r^2 \rangle_{\text{NCSM}}$~\cite{PhysRevC.68.034305} & $\langle r^2 \rangle_{\text{GFMC}}$~\cite{Pieper:2007ax}& $\langle r^2 \rangle_{so}$\\
\hline
$^7$Be &6.294(0)&5.83&6.67&0.02
\end{tabular}
\end{ruledtabular}
\end{table}

\section{Helium Halo Nuclei}

There has been a considerable increase of interest in the properties of halonuclei, as evidenced by the number of recent papers written on the topic~\cite{PhysRevLett.99.252501,PhysRevLett.102.062503,PhysRevC.71.057301,PhysRevLett.96.033002}. In $^8$He, the neutrons are assumed to completely fill all 4 states of the 1p$_{3/2}$ orbital. The calculated change in the mean-squared charge radius due to this effect is then
\begin{equation}
\langle r^2 \rangle_{so} = \sum_{m}\int  r^2\rho_{\frac{3}{2}1m} d^3r =  -0.17~\text{fm}^2\ .
\end{equation}
This shows that the contribution of a closed subshell can be significant. It is interesting that while the charge density for different projections appear to be different, the correction to the charge radius is independent of projection for any fixed $\kappa$. 

The spin-orbit term contributes a decrease in the charge radius from $^6$He to $^8$He, in addition to that predicted by the \textit{point-proton} model. The magnitude of this effect can be much larger than that of the Darwin-Foldy term in the relation (\ref{eq:rmschargeradius}), which justifies the inclusion of this spin-orbit effect. Table \ref{tab:hecomp} shows current available theoretical (no-core shell model (NCSM)~\cite{PhysRevC.73.021302} and Green's function Monte Carlo (GFMC)~\cite{Pieper:2007ax}) and experimental data, and the estimated size of the spin-orbit effect. We note that the calculated spin-orbit contribution is a maximal estimate since configuration mixing may reduce the effect.

\begin{table}
\caption{\label{tab:hecomp}Comparison of $^{6,8}$He experimental and theoretical charge radii in fm$^2$.}
\begin{ruledtabular}
\begin{tabular}{ccccc}
&$\langle r^2 \rangle_{exp}$ & $\langle r^2 \rangle_{\text{NCSM}}$\cite{PhysRevC.73.021302} & $\langle r^2 \rangle_{\text{GFMC}}$ \cite{Pieper:2007ax}& $\langle r^2 \rangle_{so}$\\
\hline
$^6$He &4.277(0)~\cite{PhysRevLett.93.142501}
&4.04&4.15&-0.08\\
$^8$He &3.722(0)~\cite{PhysRevLett.99.252501}
&3.89&3.65&-0.17
\end{tabular}
\end{ruledtabular}
\end{table}

\section{Lithium Nuclei}

It is easy to extend our results from $^8$He to $^9$Li because in the shell model the neutrons fill the same states. The unpaired $1p_{3/2}$ proton in the lithium nucleus then provides the following correction to the charge radius
\begin{equation}
\langle r^2 \rangle_{so,p} = 0.03~\text{fm}^2\ .
\end{equation}
Experimental results~\cite{PhysRevLett.96.033002} for $^9$Li and $^{11}$Li show that $^{11}$Li has the bigger charge radius. For neutrons, our calculated spin-orbit contribution is negative for the $^9$Li nucleus. Because the neutrons form a closed $p$-shell in $^{11}$Li, their spin-orbit contribution in this case is zero. Table \ref{tab:licomp} also shows that the size of the spread in theoretical data points is much larger than the spin-orbit correction, therefore it is not possible at this moment to comment on the effect that the inclusion of this term has on the accuracy of theoretical results. However, the size of this effect can be larger than existing corrections that are taken into consideration, and therefore should be accounted for as well. 

\begin{table*}
\caption{\label{tab:licomp}Comparison of $^{6,8,9,11}$Li experimental and theoretical charge radii in fm$^2$.}
\begin{ruledtabular}
\begin{tabular}{ccccccc}
&$\langle r^2 \rangle_{exp}$ & $\langle r^2 \rangle_{\text{GFMC}}$~\cite{ISI2001,PhysRevC.66.044310} & $\langle r^2 \rangle_{\text{\text{SVMC}}}$~\cite{PhysRevC.52.3013,PhysRevC.66.041302}& $\langle r^2 \rangle_{\text{DCM}}$~\cite{Tomaselli2001298}& $\langle r^2 \rangle_{\text{NCSM}}$~\cite{PhysRevC.57.3119,PhysRevC.68.034305} & $\langle r^2 \rangle_{so}$\\
\hline
$^6$Li &6.336(7)~\cite{DeJager1974479}&6.40&&7.19&5.08&-0.00\\
$^7$Li &5.726(4)~\cite{DeJager1974479}&5.71&5.80&6.46&4.83&-0.03\\
$^8$Li &5.286(8)~\cite{PhysRevLett.93.113002}&4.98&5.36&&4.38&-0.06\\
$^9$Li &4.916(6)~\cite{PhysRevLett.96.033002}&5.37&4.98&6.43&4.36&-0.09\\
$^{11}$Li &6.087(9)~\cite{PhysRevLett.96.033002}&&6.40/5.11\footnote{The second value assumes a frozen Li core.}&7.36&4.44&0.03
\end{tabular}
\end{ruledtabular}
\end{table*}

\begin{table}
\caption{\label{tab:finalcomp}Comparison of experimental charge radii with current GFMC calculations (the value for $^{11}$Li is from NCSM). The second column represents current calculations with the spin-orbit effect we calculated taken into account. All values are in fm$^2$.}
\begin{ruledtabular}
\begin{tabular}{cccc}
&$\langle r^2 \rangle_\text{GFMC}$ & $\langle r^2 \rangle_{\text{total}}$ & $\langle r^2 \rangle_{exp}$\\
\hline
$^6$He&4.15~\cite{Pieper:2007ax}&4.07&4.277(0)~\cite{PhysRevLett.93.142501}\\
$^8$He&3.65~\cite{Pieper:2007ax}&3.48&3.722(0)~\cite{PhysRevLett.99.252501}\\
$^6$Li &6.40~\cite{ISI2001,PhysRevC.66.044310}&6.40&6.336(7)~\cite{DeJager1974479}\\
$^7$Li &5.71~\cite{ISI2001,PhysRevC.66.044310}&5.68&5.726(4)~\cite{DeJager1974479}\\
$^8$Li &4.98~\cite{ISI2001,PhysRevC.66.044310}&4.92&5.286(8)~\cite{PhysRevLett.93.113002}\\
$^9$Li &5.37~\cite{ISI2001,PhysRevC.66.044310}&5.28&4.916(6)~\cite{PhysRevLett.96.033002}\\
$^{11}$Li &4.44~\cite{PhysRevC.57.3119,PhysRevC.68.034305}&4.47&6.087(9)~\cite{PhysRevLett.96.033002}\\
$^7$Be&6.67~\cite{Pieper:2007ax}&6.69&6.294(0)~\cite{PhysRevLett.102.062503}
\end{tabular}
\end{ruledtabular}
\end{table}

\section{Conclusion}

We have calculated the spin-orbit contribution to the charge density of nucleons. In $^7$Be, the spin-orbit charge density contributes approximately 2\% to the total charge density at the centre of the nucleus. The effect of the spin-orbit density on the mean-square charge radius has been shown to be independent of the radial wavefunctions, and hence the form of the (spherical) potential. A summary of the results of our calculations for the mean-square charge radius can be found in Table \ref{tab:finalcomp}.

In halonuclei such as $^8$He and $^9$Li, our results show that this effect presents a considerable contribution to the charge radius, sometimes larger than the Darwin-Foldy term and comparable to the finite size effects that are already taken into account. It is worth noting, however, that our calculations do not include configuration mixing which may reduce the size of the effect. We suggest that the spin-orbit contribution should be calculated in the same scheme as the \textit{point-proton} radii.

\appendix*
\section{}

Here we prove that the contribution of the spin-orbit charge density to the mean-square charge radius of a nucleus due to a nucleon in a spherically symmetric nuclear potential is given by \Eref{eq:onenucleonr2correction}\ . We start with the nonrelativistic expression \eref{eq:compactchargedensity}:
\[
\rho = \frac{\mu}{4M^2}\, 2i\,\boldsymbol{\sigma}_{\alpha\beta} \cdot \left[ \left( \nabla\Phi_\alpha^\dagger \right) \times \left(\nabla\Phi_\beta \right) \right]
\]
where there is an implied summation over $\alpha$, $\beta$.
This equation can be derived from \eref{eq:jzero} by substituting the nonrelativistic form of the lower component of the wavefunction \eref{eq:nonrelpsi}. Assuming a shell model allows us to write
\begin{align}
\label{eq:phidef}
\Phi &= f(r)\, \Omega_{jlm}(\theta,\phi)\ ,\\
\label{eq:omegadef}
\Omega_{jlm} &= \left( \begin{array}{c}
	C_{l\,m-\half,\,\half\,\half}^{j\,m} Y_{l\,m-\half} \\
	C_{l\,m+\half,\,\half\,-\half}^{j\,m} Y_{l\,m+\half} \end{array} \right)
\end{align}
where $f(r)$ is the (real) radial wavefunction, $\Omega_{jlm}(\theta,\phi)$ is the spherical harmonic spinor, and the $C$ are Clebsch-Gordan coefficients relating the orbital angular momentum $l$ and the spin to the total angular momentum $j$. The gradient of this wavefunction is then
\[
\grad \Phi = \Omega\,f_r\,\unitr + f\,\grad\Omega\ .
\]
where $f_r=df/dr$ and we have dropped the subscripts from $\Omega$. Substituting into \eref{eq:compactchargedensity} gives
\begin{align}
\rho = \frac{\mu}{4M^2}\, 2i\,\boldsymbol{\sigma}_{\alpha\beta} \cdot \bigg[ 
 & \frac{f f_r}{r}\Omega_\alpha^\dagger (\mathbf{r}\times\grad\Omega_\beta)
    -\frac{f f_r}{r}(\mathbf{r}\times\grad\Omega_\alpha)^\dagger \Omega_\beta
    \nonumber \\
 & + f^2\,\grad\Omega_\alpha^\dagger \times \grad\Omega_\beta \bigg]
 	\label{eq:step1}\ .
\end{align}
We split this into two parts which we treat separately: $\rho = \rho_1 + \rho_2$, the first line and second lines of \eref{eq:step1}, respectively. Using the relationship
\[
	\mathbf{\hat l}\,\Phi = (\mathbf{r}\times\mathbf{p})\Phi = -i f\,\mathbf{r}\times\grad\Omega
\]
the first line of \eref{eq:step1} can be written as
\begin{align}
\rho_1 &= \frac{2\mu}{4M^2}\bigg[
	-\frac{f_r}{r}\Omega^\dagger\ \boldsymbol{\sigma}\cdot\mathbf{\hat l}\,f\,\Omega
	-\frac{f_r}{r}(\boldsymbol{\sigma}\cdot\mathbf{\hat l}\,f\Omega)^\dagger\,\Omega
	\bigg] \nonumber \\
&= \frac{2\mu}{4M^2}\bigg[ -\frac{2 f f_r}{r} (\hat j^2 - \hat l^2 - \hat s^2)\,
    \Omega^\dagger \Omega \bigg] \nonumber \\
&= \frac{\mu}{M^2} (\kappa + 1) \frac{f f_r}{r} \Omega^\dagger \Omega
\end{align}
where $\hat j^2 = j(j+1)$ and similar for $l$ and $s$. The contribution of $\rho$ to the mean-square charge radius is given by \eref{eq:definitionofr2}. Therefore the contribution of $\rho_1$ is
\begin{align}
\langle r^2 \rangle_1 &= \frac{\mu}{M^2} (\kappa + 1) \int \frac{f f_r}{r}\,
    \Omega^\dagger \Omega\ r^2 d^3\mathbf{r} \nonumber \\
&= \frac{\mu}{M^2} (\kappa + 1) \int \frac{1}{2}\frac{d\,f^2}{dr} r^3 dr \nonumber\\
\label{eq:r1}
&= -\frac{3\mu}{2 M^2} (\kappa + 1)
\end{align}
where integration by parts was used in the last step and $f(r)\Omega(\theta,\phi)$ is assumed to be normalised to unity.

The second line of \eref{eq:step1} poses more of a challenge. Using the identity
\begin{equation*}
\grad\Omega(\theta,\phi)
 = \frac{1}{r}\frac{\partial\Omega}{\partial\theta}\,\unittheta
   + \frac{1}{r\sin\theta}\frac{\partial\Omega}{\partial\phi}\,\unitphi
\end{equation*}
where ($\unitr$, $\unittheta$, $\unitphi$) form an right-handed orthogonal basis, we write the contribution of $\rho_2$ to the mean-square charge radius as
\begin{align*}
\langle r^2 \rangle_2 &= \frac{\mu}{2M^2}\,\int i\,f^2\,\boldsymbol{\sigma}_{\alpha\beta}
    \cdot (\grad\Omega_\alpha^\dagger \times \grad\Omega_\beta)\,r^2\,d^3\mathbf{r} \\
&= \frac{\mu}{2M^2} \int i\,\boldsymbol{\sigma}_{\alpha\beta} \cdot \left(
   \frac{\partial\Omega_\alpha^\dagger}{\partial\theta}\frac{\partial\Omega_\beta}{\partial\phi}
 - \frac{\partial\Omega_\alpha^\dagger}{\partial\phi}\frac{\partial\Omega_\beta}{\partial\theta} \right)\unitr\, d\theta d\phi
\end{align*}
The Pauli matrices can be projected onto the spherical basis vectors: $\boldsymbol{\sigma} = \sigma_r\unitr + \sigma_\theta\unittheta + \sigma_\phi\unitphi$ with
\begin{subequations}
\begin{align}
\sigma_r
&= \left( \begin{array}{cc} \cos\theta & \sin\theta\,e^{-i\phi} \\ \sin\theta\,e^{i\phi} & -\cos\theta \end{array} \right) \\
\sigma_\theta
&= \left( \begin{array}{cc} -\sin\theta & \cos\theta\,e^{-i\phi} \\ \cos\theta\,e^{i\phi} & \sin\theta \end{array} \right) \\
\sigma_\phi
&= \left( \begin{array}{cc} 0 & -i\,e^{-i\phi} \\ i\,e^{i\phi} & 0 \end{array} \right)
\end{align}
\end{subequations}
Substituting $\sigma_r$ and summing over the indices $\alpha$ and $\beta$, we can write
\begin{gather}
\begin{split}
\langle r^2 \rangle_2 = \frac{\mu}{2M^2} \int i \bigg[ 
 &  \cos\theta \left( \frac{\partial\Omega_1^\dagger}{\partial\theta}\frac{\partial\Omega_1}{\partial\phi} - \frac{\partial\Omega_1^\dagger}{\partial\phi}\frac{\partial\Omega_1}{\partial\theta} \right)\\
 &+ \sin\theta\,e^{-i\phi} \left( \frac{\partial\Omega_1^\dagger}{\partial\theta}\frac{\partial\Omega_2}{\partial\phi} - \frac{\partial\Omega_1^\dagger}{\partial\phi}\frac{\partial\Omega_2}{\partial\theta} \right)\\
 &+ \sin\theta\,e^{i\phi} \left( \frac{\partial\Omega_2^\dagger}{\partial\theta}\frac{\partial\Omega_1}{\partial\phi} - \frac{\partial\Omega_2^\dagger}{\partial\phi}\frac{\partial\Omega_1}{\partial\theta} \right)\\
 &- \cos\theta \left( \frac{\partial\Omega_2^\dagger}{\partial\theta}\frac{\partial\Omega_2}{\partial\phi} - \frac{\partial\Omega_2^\dagger}{\partial\phi}\frac{\partial\Omega_2}{\partial\theta} \right)
    \bigg] d\theta\,d\phi \nonumber
\end{split}\\
\label{eq:integrals}
\langle r^2 \rangle_2 \equiv \frac{\mu}{2M^2} \left( I_1 + I_2 + I_3 + I_4\right)
\end{gather}
where $\Omega_1$ and $\Omega_2$ refer to the upper and lower components of $\Omega$, respectively, and the four integrals will be considered separately. The spherical harmonic functions in the definition of $\Omega$ \eref{eq:omegadef} can be expressed in terms of associated Legendre polynomials, $P^m_l$:
\[
Y_{l\,m} = \sqrt{\frac{2l+1}{4\pi}\frac{(l-m)!}{(l+m)!}}\, P^m_l(\cos\theta) e^{im\phi}
\]
which allows us to write $\Omega$ as
\begin{equation}
\Omega = \left( \begin{array}{c} c_1 P_l^{m-\half}(\cos\theta)\,e^{i(m-\half)\phi} \\
    c_2 P_l^{m+\half}(\cos\theta)\,e^{i(m+\half)\phi} \end{array} \right) \\
\end{equation}
where $c_1$ and $c_2$ are real coefficients. Using this definition of $\Omega$ and the transformation $u = \cos\theta$, we can express the integrals \eref{eq:integrals} as
\begin{align*}
I_1 &= 2.2\pi\,c_1^2 (m-\half) \int_{-1}^{1} u P_l^{m-\half}(u)' P_l^{m-\half}(u) du \\
I_2 &=  2\pi\,c_1 c_2 \int_{-1}^{1}
    \bigg( (m+\half) P_l^{m-\half}(u)' P_l^{m+\half}(u) \\
 &\qquad + (m-\half) P_l^{m+\half}(u)' P_l^{m-\half}(u) \bigg) \sqrt{1-u^2}\,du\\
I_3 &= I_2\\
I_4 &= - 2.2\pi\,c_2^2 (m+\half) \int_{-1}^{1} u P_l^{m+\half}(u)' P_l^{m+\half}(u) du
\end{align*}
where $P_l^m(u)'$ refer to derivatives of the associated Legendre polynomials.

$I_1$ and $I_4$ can be solved using integration by parts and the orthogonality relation
\begin{equation}
\label{eq:legendreOrtho}
\int_{-1}^{1} P_k^m(u) P_l^m(u) du = \frac{2(l+m)!}{(2l+1)(l-m)!} \delta_{k,l}\ .
\end{equation}
$I_2$ and $I_3$ are equal to each other and can be solved by transforming the derivatives of the polynomials using the identities
\begin{multline}
(u^2-1){P_{l}^{m-\half}}(u)' = \sqrt{1-u^2}P_{l}^{m+\half}(u) \\
     + (m-\nhalf)uP_{l}^{m-\half}(u) \nonumber
\end{multline}
\begin{multline}
(u^2-1){P_{l}^{m+\half}}(u)' = -(l+m+\nhalf)(l-m+\nhalf)\\ \cdot\sqrt{1-u^2}P_{l}^{m-\half}(u) - (m+\nhalf) u P_{l}^{m+\half}(u) \nonumber
\end{multline}
and then using the orthogonality relations \eref{eq:legendreOrtho}. The results are
\begin{align*}
I_1 &= - (m-\nhalf)\,(C_{l\,m-\half,\half\,\half}^{j\,m})^2\\
I_2 &= - C_{l\,m-\half,\half\,\half}^{j\,m}C_{l\,m+\half,\half\,-\half}^{j\,m}
    \sqrt{(l-m+\nhalf)(l+m+\nhalf)}\\
    &= I_3\\
I_4 &= (m+\nhalf)\,(C_{l\,m+\half,\half\,-\half}^{j\,m})^2\ .
\end{align*}
Substituting into \eref{eq:integrals} and evaluating the Clebsch-Gordan coefficients we obtain
\begin{equation}
\langle r^2 \rangle_2 = \frac{\mu}{2M^2}(\kappa+1)\ .
\end{equation}
Adding this contribution to \eref{eq:r1} gives the total change in mean-square charge radius due to a single nucleon
\begin{equation}
\langle r^2 \rangle_{so}^{(1)} = \langle r^2 \rangle_1 + \langle r^2 \rangle_2
    = - \frac{\mu}{M^2} (\kappa + 1)\ .
\end{equation}
The contribution of a filled subshell to the normalised mean-square charge radius is therefore
\[
\frac{1}{Z}\sum_\textrm{subshell} \langle r^2 \rangle_{so}^{(1)} 
= -\frac{2j+1}{Z} \frac{\mu}{M^2} (\kappa + 1)
\]
in agreement with \cite{Bertozzi1972408}. For a spin-orbit doublet, on the other hand, we find
\[
\frac{1}{Z}\sum_\textrm{so doublet} \langle r^2 \rangle_{so}^{(1)} 
= \frac{2(l+1)}{Z} \frac{\mu}{M^2} l - \frac{2l}{Z}\frac{\mu}{M^2} (l+1) = 0\ .
\]

\acknowledgments
The authors thank R. Wiringa for useful discussions. This work is supported by the Australian Research Council.

\bibliography{cdref}

\begin{thebibliography}{10}%
\makeatletter
\providecommand \@ifxundefined [1]{%
 \ifx #1\undefined \expandafter \@firstoftwo
 \else \expandafter \@secondoftwo
\fi
}%
\providecommand \@ifnum [1]{%
 \ifnum #1\expandafter \@firstoftwo
 \else \expandafter \@secondoftwo
\fi
}%
\providecommand \enquote [1]{``#1''}%
\providecommand \bibnamefont  [1]{#1}%
\providecommand \bibfnamefont [1]{#1}%
\providecommand \citenamefont [1]{#1}%
\providecommand\href[0]{\@sanitize\@href}%
\providecommand\@href[1]{\endgroup\@@startlink{#1}\endgroup\@@href}%
\providecommand\@@href[1]{#1\@@endlink}%
\providecommand \@sanitize [0]{\begingroup\catcode`\&12\catcode`\#12\relax}%
\@ifxundefined \pdfoutput {\@firstoftwo}{%
 \@ifnum{\z@=\pdfoutput}{\@firstoftwo}{\@secondoftwo}%
}{%
 \providecommand\@@startlink[1]{\leavevmode\special{html:<a href="#1">}}%
 \providecommand\@@endlink[0]{\special{html:</a>}}%
}{%
 \providecommand\@@startlink[1]{%
  \leavevmode
  \pdfstartlink
   attr{/Border[0 0 1 ]/H/I/C[0 1 1]}%
   user{/Subtype/Link/A<</Type/Action/S/URI/URI(#1)>>}%
  \relax
 }%
 \providecommand\@@endlink[0]{\pdfendlink}%
}%
\providecommand \url  [0]{\begingroup\@sanitize \@url }%
\providecommand \@url [1]{\endgroup\@href {#1}{\urlprefix}}%
\providecommand \urlprefix [0]{URL }%
\providecommand \Eprint[0]{\href }%
\@ifxundefined \urlstyle {%
  \providecommand \doi [1]{doi:\discretionary{}{}{}#1}%
}{%
  \providecommand \doi [0]{doi:\discretionary{}{}{}\begingroup
  \urlstyle{rm}\Url }%
}%
\providecommand \doibase [0]{http://dx.doi.org/}%
\providecommand \Doi[1]{\href{\doibase#1}}%
\providecommand \bibAnnote [3]{%
  \BibitemShut{#1}%
  \begin{quotation}\noindent
    \textsc{Key:}\ #2\\\textsc{Annotation:}\ #3%
  \end{quotation}%
}%
\providecommand \bibAnnoteFile [2]{%
  \IfFileExists{#2}{\bibAnnote {#1} {#2} {\input{#2}}}{}%
}%
\providecommand \typeout [0]{\immediate \write \m@ne }%
\providecommand \selectlanguage [0]{\@gobble}%
\providecommand \bibinfo [0]{\@secondoftwo}%
\providecommand \bibfield [0]{\@secondoftwo}%
\providecommand \translation [1]{[#1]}%
\providecommand \BibitemOpen[0]{}%
\providecommand \bibitemStop [0]{}%
\providecommand \bibitemNoStop [0]{.\EOS\space}%
\providecommand \EOS [0]{\spacefactor3000\relax}%
\providecommand \BibitemShut [1]{\csname bibitem#1\endcsname}%
\bibitem{Bertozzi1972408}%
  \BibitemOpen
  \bibfield{author}{%
  \bibinfo {author} {\bibfnamefont{W.}~\bibnamefont{Bertozzi}}, \bibinfo
  {author} {\bibfnamefont{J.}~\bibnamefont{Friar}}, \bibinfo {author}
  {\bibfnamefont{J.}~\bibnamefont{Heisenberg}},\ and\ \bibinfo {author}
  {\bibfnamefont{J.~W.}\ \bibnamefont{Negele}},\ }%
  \bibfield{journal}{%
  \Doi{DOI: 10.1016/0370-2693(72)90662-4}{\bibinfo {journal} {Physics Letters
  B}}\ }%
  \textbf{\bibinfo {volume} {41}},\ \bibinfo {pages} {408 } (\bibinfo {year}
  {1972})%
  \bibAnnoteFile{NoStop}{Bertozzi1972408}%
\bibitem{Kurasawa:2000ve}%
  \BibitemOpen
  \bibfield{author}{%
  \bibinfo {author} {\bibfnamefont{H.}~\bibnamefont{Kurasawa}}\ and\ \bibinfo
  {author} {\bibfnamefont{T.}~\bibnamefont{Suzuki}},\ }%
  \bibfield{journal}{%
  \Doi{10.1103/PhysRevC.62.054303}{\bibinfo {journal} {Phys. Rev. C}}\ }%
  \textbf{\bibinfo {volume} {62}},\ \bibinfo {pages} {054303} (\bibinfo {year}
  {2000})%
  \bibAnnoteFile{NoStop}{Kurasawa:2000ve}%
\bibitem{PhysRevC.13.245}%
  \BibitemOpen
  \bibfield{author}{%
  \bibinfo {author} {\bibfnamefont{H.}~\bibnamefont{Chandra}}\ and\ \bibinfo
  {author} {\bibfnamefont{G.}~\bibnamefont{Sauer}},\ }%
  \bibfield{journal}{%
  \Doi{10.1103/PhysRevC.13.245}{\bibinfo {journal} {Phys. Rev. C}}\ }%
  \textbf{\bibinfo {volume} {13}},\ \bibinfo {pages} {245} (\bibinfo {month}
  {Jan}\ \bibinfo {year} {1976})%
  \bibAnnoteFile{NoStop}{PhysRevC.13.245}%
\bibitem{PhysRev.78.29}%
  \BibitemOpen
  \bibfield{author}{%
  \bibinfo {author} {\bibfnamefont{L.~L.}\ \bibnamefont{Foldy}}\ and\ \bibinfo
  {author} {\bibfnamefont{S.~A.}\ \bibnamefont{Wouthuysen}},\ }%
  \bibfield{journal}{%
  \Doi{10.1103/PhysRev.78.29}{\bibinfo {journal} {Phys. Rev.}}\ }%
  \textbf{\bibinfo {volume} {78}},\ \bibinfo {pages} {29} (\bibinfo {month}
  {Apr}\ \bibinfo {year} {1950})%
  \bibAnnoteFile{NoStop}{PhysRev.78.29}%
\bibitem{bjorkendrell1964}%
  \BibitemOpen
  \bibfield{author}{%
  \bibinfo {author} {\bibfnamefont{J.~D.}\ \bibnamefont{Bjorken}}\ and\
  \bibinfo {author} {\bibfnamefont{S.~D.}\ \bibnamefont{Drell}},\ }%
  \emph{\bibinfo {title} {Relativistic Quantum Mechanics}}\ (\bibinfo
  {publisher} {McGraw-Hill, New York},\ \bibinfo {year} {1964})%
  \bibAnnoteFile{NoStop}{bjorkendrell1964}%
\bibitem{PhysRevA.56.4579}%
  \BibitemOpen
  \bibfield{author}{%
  \bibinfo {author} {\bibfnamefont{J.~L.}\ \bibnamefont{Friar}}, \bibinfo
  {author} {\bibfnamefont{J.}~\bibnamefont{Martorell}},\ and\ \bibinfo {author}
  {\bibfnamefont{D.~W.~L.}\ \bibnamefont{Sprung}},\ }%
  \bibfield{journal}{%
  \Doi{10.1103/PhysRevA.56.4579}{\bibinfo {journal} {Phys. Rev. A}}\ }%
  \textbf{\bibinfo {volume} {56}},\ \bibinfo {pages} {4579} (\bibinfo {month}
  {Dec}\ \bibinfo {year} {1997})%
  \bibAnnoteFile{NoStop}{PhysRevA.56.4579}%
\bibitem{PhysRevC.73.021302}%
  \BibitemOpen
  \bibfield{author}{%
  \bibinfo {author} {\bibfnamefont{E.}~\bibnamefont{Caurier}}\ and\ \bibinfo
  {author} {\bibfnamefont{P.}~\bibnamefont{Navr\'atil}},\ }%
  \bibfield{journal}{%
  \Doi{10.1103/PhysRevC.73.021302}{\bibinfo {journal} {Phys. Rev. C}}\ }%
  \textbf{\bibinfo {volume} {73}},\ \bibinfo {pages} {021302} (\bibinfo {month}
  {Feb}\ \bibinfo {year} {2006})%
  \bibAnnoteFile{NoStop}{PhysRevC.73.021302}%
\bibitem{Pieper:2007ax}%
  \BibitemOpen
  \bibfield{author}{%
  \bibinfo {author} {\bibfnamefont{S.~C.}\ \bibnamefont{Pieper}},\ }%
  \bibfield{journal}{%
  \Doi{10.1393/ncr/i2009-10039-1}{\bibinfo {journal} {Riv. Nuovo Cim.}}\ }%
  \textbf{\bibinfo {volume} {031}},\ \bibinfo {pages} {709} (\bibinfo {year}
  {2008})%
  \bibAnnoteFile{NoStop}{Pieper:2007ax}%
\bibitem{ISI2001}%
  \BibitemOpen
  \bibfield{author}{%
  \bibinfo {author} {\bibfnamefont{S.}~\bibnamefont{Pieper}}\ and\ \bibinfo
  {author} {\bibfnamefont{R.}~\bibnamefont{Wiringa}},\ }%
  \bibfield{journal}{%
  \bibinfo {journal} {Annual Review of Nuclear and Particle Science}\ }%
  \textbf{\bibinfo {volume} {51}},\ \bibinfo {pages} {53} (\bibinfo {year}
  {2001})%
  \bibAnnoteFile{NoStop}{ISI2001}%
\bibitem{PhysRevC.66.044310}%
  \BibitemOpen
  \bibfield{author}{%
  \bibinfo {author} {\bibfnamefont{S.~C.}\ \bibnamefont{Pieper}}, \bibinfo
  {author} {\bibfnamefont{K.}~\bibnamefont{Varga}},\ and\ \bibinfo {author}
  {\bibfnamefont{R.~B.}\ \bibnamefont{Wiringa}},\ }%
  \bibfield{journal}{%
  \Doi{10.1103/PhysRevC.66.044310}{\bibinfo {journal} {Phys. Rev. C}}\ }%
  \textbf{\bibinfo {volume} {66}},\ \bibinfo {pages} {044310} (\bibinfo {month}
  {Oct}\ \bibinfo {year} {2002})%
  \bibAnnoteFile{NoStop}{PhysRevC.66.044310}%
\bibitem{PhysRevC.52.3013}%
  \BibitemOpen
  \bibfield{author}{%
  \bibinfo {author} {\bibfnamefont{K.}~\bibnamefont{Varga}}, \bibinfo {author}
  {\bibfnamefont{Y.}~\bibnamefont{Suzuki}},\ and\ \bibinfo {author}
  {\bibfnamefont{I.}~\bibnamefont{Tanihata}},\ }%
  \bibfield{journal}{%
  \Doi{10.1103/PhysRevC.52.3013}{\bibinfo {journal} {Phys. Rev. C}}\ }%
  \textbf{\bibinfo {volume} {52}},\ \bibinfo {pages} {3013} (\bibinfo {month}
  {Dec}\ \bibinfo {year} {1995})%
  \bibAnnoteFile{NoStop}{PhysRevC.52.3013}%
\bibitem{PhysRevC.66.041302}%
  \BibitemOpen
  \bibfield{author}{%
  \bibinfo {author} {\bibfnamefont{K.}~\bibnamefont{Varga}}, \bibinfo {author}
  {\bibfnamefont{Y.}~\bibnamefont{Suzuki}},\ and\ \bibinfo {author}
  {\bibfnamefont{R.~G.}\ \bibnamefont{Lovas}},\ }%
  \bibfield{journal}{%
  \Doi{10.1103/PhysRevC.66.041302}{\bibinfo {journal} {Phys. Rev. C}}\ }%
  \textbf{\bibinfo {volume} {66}},\ \bibinfo {pages} {041302} (\bibinfo {month}
  {Oct}\ \bibinfo {year} {2002})%
  \bibAnnoteFile{NoStop}{PhysRevC.66.041302}%
\bibitem{Tomaselli2001298}%
  \BibitemOpen
  \bibfield{author}{%
  \bibinfo {author} {\bibfnamefont{M.}~\bibnamefont{Tomaselli}}, \bibinfo
  {author} {\bibfnamefont{S.}~\bibnamefont{Fritzsche}}, \bibinfo {author}
  {\bibfnamefont{A.}~\bibnamefont{Dax}}, \bibinfo {author}
  {\bibfnamefont{P.}~\bibnamefont{Egelhof}}, \bibinfo {author}
  {\bibfnamefont{C.}~\bibnamefont{Kozhuharov}}, \bibinfo {author}
  {\bibfnamefont{T.}~\bibnamefont{K\"uhl}}, \bibinfo {author}
  {\bibfnamefont{D.}~\bibnamefont{Marx}}, \bibinfo {author}
  {\bibfnamefont{M.}~\bibnamefont{Mutterer}}, \bibinfo {author}
  {\bibfnamefont{S.~R.}\ \bibnamefont{Neumaier}}, \bibinfo {author}
  {\bibfnamefont{W.}~\bibnamefont{N\"ortersh\"auser}}, \bibinfo {author}
  {\bibfnamefont{H.}~\bibnamefont{Wang}},\ and\ \bibinfo {author}
  {\bibfnamefont{H.~J.}\ \bibnamefont{Kluge}},\ }%
  \bibfield{journal}{%
  \Doi{DOI: 10.1016/S0375-9474(01)00963-0}{\bibinfo {journal} {Nuclear Physics
  A}}\ }%
  \textbf{\bibinfo {volume} {690}},\ \bibinfo {pages} {298 } (\bibinfo {year}
  {2001})%
  \bibAnnoteFile{NoStop}{Tomaselli2001298}%
\bibitem{PhysRevC.57.3119}%
  \BibitemOpen
  \bibfield{author}{%
  \bibinfo {author} {\bibfnamefont{P.}~\bibnamefont{Navr\'atil}}\ and\ \bibinfo
  {author} {\bibfnamefont{B.~R.}\ \bibnamefont{Barrett}},\ }%
  \bibfield{journal}{%
  \Doi{10.1103/PhysRevC.57.3119}{\bibinfo {journal} {Phys. Rev. C}}\ }%
  \textbf{\bibinfo {volume} {57}},\ \bibinfo {pages} {3119} (\bibinfo {month}
  {Jun}\ \bibinfo {year} {1998})%
  \bibAnnoteFile{NoStop}{PhysRevC.57.3119}%
\bibitem{PhysRevC.68.034305}%
  \BibitemOpen
  \bibfield{author}{%
  \bibinfo {author} {\bibfnamefont{P.}~\bibnamefont{Navr\'atil}}\ and\ \bibinfo
  {author} {\bibfnamefont{W.~E.}\ \bibnamefont{Ormand}},\ }%
  \bibfield{journal}{%
  \Doi{10.1103/PhysRevC.68.034305}{\bibinfo {journal} {Phys. Rev. C}}\ }%
  \textbf{\bibinfo {volume} {68}},\ \bibinfo {pages} {034305} (\bibinfo {month}
  {Sep}\ \bibinfo {year} {2003})%
  \bibAnnoteFile{NoStop}{PhysRevC.68.034305}%
\bibitem{landau1982}%
  \BibitemOpen
  \bibfield{author}{%
  \bibinfo {author} {\bibfnamefont{E.~L.}\ \bibnamefont{V~B~Berestetskii},
  \bibfnamefont{L.~P.~Pitaevskii}},\ }%
  \emph{\bibinfo {title} {Quantum Electrodynamics : Volume 4}}\ (\bibinfo
  {publisher} {Twayne Publishers},\ \bibinfo {address} {Boston},\ \bibinfo
  {year} {1982})%
  \bibAnnoteFile{NoStop}{landau1982}%
\bibitem{PDBook}%
  \BibitemOpen
  \bibfield{author}{%
  \bibinfo {author} {\bibfnamefont{W.-M.}\ \bibnamefont{Yao}},\ }%
  \bibfield{journal}{%
  \bibinfo {journal} {{J. Phys. G}}\ }%
  \textbf{\bibinfo {volume} {33}} (\bibinfo {year} {2006})%
  \bibAnnoteFile{NoStop}{PDBook}%
\bibitem{PhysRevLett.96.033002}%
  \BibitemOpen
  \bibfield{author}{%
  \bibinfo {author} {\bibfnamefont{R.}~\bibnamefont{S\'anchez}}, \bibinfo
  {author} {\bibfnamefont{W.}~\bibnamefont{N\"ortersh\"auser}}, \bibinfo
  {author} {\bibfnamefont{G.}~\bibnamefont{Ewald}}, \bibinfo {author}
  {\bibfnamefont{D.}~\bibnamefont{Albers}}, \bibinfo {author}
  {\bibfnamefont{J.}~\bibnamefont{Behr}}, \bibinfo {author}
  {\bibfnamefont{P.}~\bibnamefont{Bricault}}, \bibinfo {author}
  {\bibfnamefont{B.~A.}\ \bibnamefont{Bushaw}}, \bibinfo {author}
  {\bibfnamefont{A.}~\bibnamefont{Dax}}, \bibinfo {author}
  {\bibfnamefont{J.}~\bibnamefont{Dilling}}, \bibinfo {author}
  {\bibfnamefont{M.}~\bibnamefont{Dombsky}}, \bibinfo {author}
  {\bibfnamefont{G.~W.~F.}\ \bibnamefont{Drake}}, \bibinfo {author}
  {\bibfnamefont{S.}~\bibnamefont{G\"otte}}, \bibinfo {author}
  {\bibfnamefont{R.}~\bibnamefont{Kirchner}}, \bibinfo {author}
  {\bibfnamefont{H.-J.}\ \bibnamefont{Kluge}}, \bibinfo {author}
  {\bibfnamefont{T.}~\bibnamefont{K\"uhl}}, \bibinfo {author}
  {\bibfnamefont{J.}~\bibnamefont{Lassen}}, \bibinfo {author}
  {\bibfnamefont{C.~D.~P.}\ \bibnamefont{Levy}}, \bibinfo {author}
  {\bibfnamefont{M.~R.}\ \bibnamefont{Pearson}}, \bibinfo {author}
  {\bibfnamefont{E.~J.}\ \bibnamefont{Prime}}, \bibinfo {author}
  {\bibfnamefont{V.}~\bibnamefont{Ryjkov}}, \bibinfo {author}
  {\bibfnamefont{A.}~\bibnamefont{Wojtaszek}}, \bibinfo {author}
  {\bibfnamefont{Z.-C.}\ \bibnamefont{Yan}},\ and\ \bibinfo {author}
  {\bibfnamefont{C.}~\bibnamefont{Zimmermann}},\ }%
  \bibfield{journal}{%
  \Doi{10.1103/PhysRevLett.96.033002}{\bibinfo {journal} {Phys. Rev. Lett.}}\
  }%
  \textbf{\bibinfo {volume} {96}},\ \bibinfo {pages} {033002} (\bibinfo {month}
  {Jan}\ \bibinfo {year} {2006})%
  \bibAnnoteFile{NoStop}{PhysRevLett.96.033002}%
\bibitem{griffiths2005}%
  \BibitemOpen
  \bibfield{author}{%
  \bibinfo {author} {\bibfnamefont{D.}~\bibnamefont{Griffiths}},\ }%
  \emph{\bibinfo {title} {Introduction to Quantum Mechanics}}\ (\bibinfo
  {publisher} {Westview},\ \bibinfo {address} {Boulder},\ \bibinfo {year}
  {2005})%
  \bibAnnoteFile{NoStop}{griffiths2005}%
\bibitem{PhysRevLett.102.062503}%
  \BibitemOpen
  \bibfield{author}{%
  \bibinfo {author} {\bibfnamefont{W.}~\bibnamefont{N\"ortersh\"auser}},
  \bibinfo {author} {\bibfnamefont{D.}~\bibnamefont{Tiedemann}}, \bibinfo
  {author} {\bibfnamefont{M.}~\bibnamefont{\ifmmode~\check{Z}\else
  \v{Z}\fi{}\'akov\'a}}, \bibinfo {author}
  {\bibfnamefont{Z.}~\bibnamefont{Andjelkovic}}, \bibinfo {author}
  {\bibfnamefont{K.}~\bibnamefont{Blaum}}, \bibinfo {author}
  {\bibfnamefont{M.~L.}\ \bibnamefont{Bissell}}, \bibinfo {author}
  {\bibfnamefont{R.}~\bibnamefont{Cazan}}, \bibinfo {author}
  {\bibfnamefont{G.~W.~F.}\ \bibnamefont{Drake}}, \bibinfo {author}
  {\bibfnamefont{C.}~\bibnamefont{Geppert}}, \bibinfo {author}
  {\bibfnamefont{M.}~\bibnamefont{Kowalska}}, \bibinfo {author}
  {\bibfnamefont{J.}~\bibnamefont{Kr\"amer}}, \bibinfo {author}
  {\bibfnamefont{A.}~\bibnamefont{Krieger}}, \bibinfo {author}
  {\bibfnamefont{R.}~\bibnamefont{Neugart}}, \bibinfo {author}
  {\bibfnamefont{R.}~\bibnamefont{S\'anchez}}, \bibinfo {author}
  {\bibfnamefont{F.}~\bibnamefont{Schmidt-Kaler}}, \bibinfo {author}
  {\bibfnamefont{Z.-C.}\ \bibnamefont{Yan}}, \bibinfo {author}
  {\bibfnamefont{D.~T.}\ \bibnamefont{Yordanov}},\ and\ \bibinfo {author}
  {\bibfnamefont{C.}~\bibnamefont{Zimmermann}},\ }%
  \bibfield{journal}{%
  \Doi{10.1103/PhysRevLett.102.062503}{\bibinfo {journal} {Phys. Rev. Lett.}}\
  }%
  \textbf{\bibinfo {volume} {102}},\ \bibinfo {pages} {062503} (\bibinfo
  {month} {Feb}\ \bibinfo {year} {2009})%
  \bibAnnoteFile{NoStop}{PhysRevLett.102.062503}%
\bibitem{PhysRevLett.99.252501}%
  \BibitemOpen
  \bibfield{author}{%
  \bibinfo {author} {\bibfnamefont{P.}~\bibnamefont{Mueller}}, \bibinfo
  {author} {\bibfnamefont{I.~A.}\ \bibnamefont{Sulai}}, \bibinfo {author}
  {\bibfnamefont{A.~C.~C.}\ \bibnamefont{Villari}}, \bibinfo {author}
  {\bibfnamefont{J.~A.}\ \bibnamefont{Alc\'antara-N\'u\~nez}}, \bibinfo
  {author} {\bibfnamefont{R.}~\bibnamefont{Alves-Cond\'e}}, \bibinfo {author}
  {\bibfnamefont{K.}~\bibnamefont{Bailey}}, \bibinfo {author}
  {\bibfnamefont{G.~W.~F.}\ \bibnamefont{Drake}}, \bibinfo {author}
  {\bibfnamefont{M.}~\bibnamefont{Dubois}}, \bibinfo {author}
  {\bibfnamefont{C.}~\bibnamefont{El\'eon}}, \bibinfo {author}
  {\bibfnamefont{G.}~\bibnamefont{Gaubert}}, \bibinfo {author}
  {\bibfnamefont{R.~J.}\ \bibnamefont{Holt}}, \bibinfo {author}
  {\bibfnamefont{R.~V.~F.}\ \bibnamefont{Janssens}}, \bibinfo {author}
  {\bibfnamefont{N.}~\bibnamefont{Lecesne}}, \bibinfo {author}
  {\bibfnamefont{Z.-T.}\ \bibnamefont{Lu}}, \bibinfo {author}
  {\bibfnamefont{T.~P.}\ \bibnamefont{O'Connor}}, \bibinfo {author}
  {\bibfnamefont{M.-G.}\ \bibnamefont{Saint-Laurent}}, \bibinfo {author}
  {\bibfnamefont{J.-C.}\ \bibnamefont{Thomas}},\ and\ \bibinfo {author}
  {\bibfnamefont{L.-B.}\ \bibnamefont{Wang}},\ }%
  \bibfield{journal}{%
  \Doi{10.1103/PhysRevLett.99.252501}{\bibinfo {journal} {Phys. Rev. Lett.}}\
  }%
  \textbf{\bibinfo {volume} {99}},\ \bibinfo {pages} {252501} (\bibinfo {month}
  {Dec}\ \bibinfo {year} {2007})%
  \bibAnnoteFile{NoStop}{PhysRevLett.99.252501}%
\bibitem{PhysRevC.71.057301}%
  \BibitemOpen
  \bibfield{author}{%
  \bibinfo {author} {\bibfnamefont{B.~V.}\ \bibnamefont{Danilin}}, \bibinfo
  {author} {\bibfnamefont{S.~N.}\ \bibnamefont{Ershov}},\ and\ \bibinfo
  {author} {\bibfnamefont{J.~S.}\ \bibnamefont{Vaagen}},\ }%
  \bibfield{journal}{%
  \Doi{10.1103/PhysRevC.71.057301}{\bibinfo {journal} {Phys. Rev. C}}\ }%
  \textbf{\bibinfo {volume} {71}},\ \bibinfo {pages} {057301} (\bibinfo {month}
  {May}\ \bibinfo {year} {2005})%
  \bibAnnoteFile{NoStop}{PhysRevC.71.057301}%
\bibitem{PhysRevLett.93.142501}%
  \BibitemOpen
  \bibfield{author}{%
  \bibinfo {author} {\bibfnamefont{L.-B.}\ \bibnamefont{Wang}}, \bibinfo
  {author} {\bibfnamefont{P.}~\bibnamefont{Mueller}}, \bibinfo {author}
  {\bibfnamefont{K.}~\bibnamefont{Bailey}}, \bibinfo {author}
  {\bibfnamefont{G.~W.~F.}\ \bibnamefont{Drake}}, \bibinfo {author}
  {\bibfnamefont{J.~P.}\ \bibnamefont{Greene}}, \bibinfo {author}
  {\bibfnamefont{D.}~\bibnamefont{Henderson}}, \bibinfo {author}
  {\bibfnamefont{R.~J.}\ \bibnamefont{Holt}}, \bibinfo {author}
  {\bibfnamefont{R.~V.~F.}\ \bibnamefont{Janssens}}, \bibinfo {author}
  {\bibfnamefont{C.~L.}\ \bibnamefont{Jiang}}, \bibinfo {author}
  {\bibfnamefont{Z.-T.}\ \bibnamefont{Lu}}, \bibinfo {author}
  {\bibfnamefont{T.~P.}\ \bibnamefont{O'Connor}}, \bibinfo {author}
  {\bibfnamefont{R.~C.}\ \bibnamefont{Pardo}}, \bibinfo {author}
  {\bibfnamefont{K.~E.}\ \bibnamefont{Rehm}}, \bibinfo {author}
  {\bibfnamefont{J.~P.}\ \bibnamefont{Schiffer}},\ and\ \bibinfo {author}
  {\bibfnamefont{X.~D.}\ \bibnamefont{Tang}},\ }%
  \bibfield{journal}{%
  \Doi{10.1103/PhysRevLett.93.142501}{\bibinfo {journal} {Phys. Rev. Lett.}}\
  }%
  \textbf{\bibinfo {volume} {93}},\ \bibinfo {pages} {142501} (\bibinfo {month}
  {Sep}\ \bibinfo {year} {2004})%
  \bibAnnoteFile{NoStop}{PhysRevLett.93.142501}%
\bibitem{DeJager1974479}%
  \BibitemOpen
  \bibfield{author}{%
  \bibinfo {author} {\bibfnamefont{C.~D.}\ \bibnamefont{Jager}}, \bibinfo
  {author} {\bibfnamefont{H.~D.}\ \bibnamefont{Vries}},\ and\ \bibinfo {author}
  {\bibfnamefont{C.~D.}\ \bibnamefont{Vries}},\ }%
  \bibfield{journal}{%
  \Doi{DOI: 10.1016/S0092-640X(74)80002-1}{\bibinfo {journal} {At. Data Nucl.
  Data Tables}}\ }%
  \textbf{\bibinfo {volume} {14}},\ \bibinfo {pages} {479 } (\bibinfo {year}
  {1974})%
  \bibAnnoteFile{NoStop}{DeJager1974479}%
\bibitem{PhysRevLett.93.113002}%
  \BibitemOpen
  \bibfield{author}{%
  \bibinfo {author} {\bibfnamefont{G.}~\bibnamefont{Ewald}}, \bibinfo {author}
  {\bibfnamefont{W.}~\bibnamefont{N\"ortersh\"auser}}, \bibinfo {author}
  {\bibfnamefont{A.}~\bibnamefont{Dax}}, \bibinfo {author}
  {\bibfnamefont{S.}~\bibnamefont{G\"otte}}, \bibinfo {author}
  {\bibfnamefont{R.}~\bibnamefont{Kirchner}}, \bibinfo {author}
  {\bibfnamefont{H.-J.}\ \bibnamefont{Kluge}}, \bibinfo {author}
  {\bibfnamefont{T.}~\bibnamefont{K\"uhl}}, \bibinfo {author}
  {\bibfnamefont{R.}~\bibnamefont{Sanchez}}, \bibinfo {author}
  {\bibfnamefont{A.}~\bibnamefont{Wojtaszek}}, \bibinfo {author}
  {\bibfnamefont{B.~A.}\ \bibnamefont{Bushaw}}, \bibinfo {author}
  {\bibfnamefont{G.~W.~F.}\ \bibnamefont{Drake}}, \bibinfo {author}
  {\bibfnamefont{Z.-C.}\ \bibnamefont{Yan}},\ and\ \bibinfo {author}
  {\bibfnamefont{C.}~\bibnamefont{Zimmermann}},\ }%
  \bibfield{journal}{%
  \Doi{10.1103/PhysRevLett.93.113002}{\bibinfo {journal} {Phys. Rev. Lett.}}\
  }%
  \textbf{\bibinfo {volume} {93}},\ \bibinfo {pages} {113002} (\bibinfo {month}
  {Sep}\ \bibinfo {year} {2004})%
  \bibAnnoteFile{NoStop}{PhysRevLett.93.113002}%
\end{thebibliography}%

\end{document}